\documentclass[pra,aps,twocolumn]{revtex4-2}
\usepackage{bm,dcolumn,amsmath,graphicx,amsfonts,amssymb,lipsum,mathtools}
\usepackage{epsfig}
\usepackage{tikz}
\usepackage{xcolor}

\usepackage{graphicx,subfigure}

\begin{document}

	\title{Hf$^{12+}$ ion: Highly Charged Ion for Next-Generation Atomic Clocks and Tests of Fundamental Physics}

	\author{Saleh O. Allehabi$^{1}$, V. A. Dzuba$^2$, and V. V. Flambaum$^2$}
	
	\affiliation{$^{1}$Department of Physics, Faculty of Science, Islamic University of Madinah, Madinah 42351, Saudi Arabia}
	\affiliation{$^2$School of Physics, University of New South Wales, Sydney 2052, Australia}

\date{\today}
	
\begin{abstract}
	
We use advanced computational techniques to study the electronic structure of the Hf$^{12+}$ ion, with the goal of assessing its potential for use in highly accurate atomic optical clocks and search for new physics. Such clocks should combine low sensitivity to external perturbations with high sensitivity to a possible time variation of the fine-structure constant $\alpha$.
The system features two clock transitions. One is an  $f-p$ transition in terms of single-electron states, which exhibits strong sensitivity to 
variations in $\alpha$. 
The other is an electric-quadrupole (E2) transition between states of the ground-state configuration, which can serve as an anchor transition for measuring one frequency against the other.

All three relevant states possess very small and nearly equal static dipole polarizabilities, resulting in an extremely small blackbody-radiation shift. The quadrupole shift is also small and can be further suppressed. Altogether, Hf$^{12+}$
 appears to be a highly promising candidate for both precision timekeeping and searches for new physics.	
	
\end{abstract}
	
\maketitle
	
\section{Introduction}

Atomic clocks have achieved an exceptional level of accuracy, enabling stringent tests of fundamental physics, major improvements in timekeeping, and significant advancements in metrology \cite{ludlow2015optical, huntemann2016single,Dimarcq2024}. Among their many applications, one of the most intriguing is the search for possible variations of the fine-structure constant, $\alpha$, which governs the strength of electromagnetic interactions \cite{uzan2011varying, PhysRevLett.74.3511}. 
Theories beyond the Standard Model, including those involving couplings to dark matter, predict that $\alpha$
may exhibit spatial or temporal variations, making high-precision spectroscopic measurements a powerful tool for testing such scenarios \cite{DzuFla00,derevianko2014hunting}.
Currently, the most stringent laboratory constraint on the temporal variation of $\alpha$ is
$\dot \alpha/\alpha=(1.8 \pm 2.5) \times 10^{-19}  \ {\rm yr}^{-1}$ 
obtained from a comparison of two optical clock transitions in 
$^{171}$Yb$^+$ \cite{Peik2023}.

It was demonstrated in earlier works~\cite{DzuFlaWeb99a} that the highest sensitivity of atomic transition frequencies to variations in the fine-structure constant, $\alpha$, occurs for single-electron $s$–$p$ or $p$–$f$ transitions. Additional enhancement arises in systems with large nuclear charge $Z$ ($\sim Z^2$) and high ionization degree $Z_i$ ( $\sim Z_i^2$). On the other hand,   systematic effects , such as the black body radiation shift, should rapidly  decrease with $Z_i$  due to small ion size \cite{BDFO12}. Highly charged ions (HCIs) are therefore particularly promising candidates~\cite{BDF-HCI10,DDF-PRL12}. However, in most HCIs, transitions between states of different electronic configurations typically lie outside the optical frequency range.

The situation changes when one takes into account that the ordering of electronic states varies with changing $Z$ or $Z_i$. Energy of higher $l$ orbitals, such $f$ orbitals, go down with increase of $Z_i$ much faster than energy of low $l$ orbitals $s,p$. This leads to so-called level crossings~\cite{BDF-HCI10,BDFO-hole11,BDFO12}, which can bring transitions between different configurations into the optical domain.

If such transitions also possess the defining characteristics of high-quality clock transitions, then combining their large $\alpha$-sensitivity with the exceptional measurement accuracy achievable in optical clocks makes them excellent candidates for probing variations of the fine-structure constant.

A number of such transitions were investigated theoretically (see, e.g. reviews ~\cite{KozlovHCIreview,Sahoo2023} and recent articles~\cite{actinides,DF-2024,ADF2024,Ir17+Liu,Wu2025,Pr9+}). Some of them were also investigated experimentally. This includes Ho$^{14}$ ions~\cite{Ho14+}, Ir$^{17+}$ ion~\cite{IrPRL15},  Ar$^{13+}$~\cite{Ar13+} and Os$^{16+}$~\cite{Os16+} ions.

In this work, we investigate the feasibility of employing the Hf$^{12+}$ ion (hafnium in a twelve-times ionized state) as a candidate for an optical atomic clock that is highly sensitive to variations in the fine-structure constant, $\alpha$. 
The level structure of ion is presented on Fig.~\ref{f:hf12} (this diagram is based on the calculations of present work, see below).

The Hf$^{12+}$ ion exhibits several advantageous electronic properties:
\begin{enumerate}
\item { The ion features two clock transitions that exhibit different sensitivities to variation of the fine-structure constant. This allows for precise monitoring of potential $\alpha$-variation through long-term comparison of the corresponding transition frequencies.}
\item Its transitions, particularly the $p$–$f$ transitions, experience strong relativistic effects, enhancing their sensitivity to variations in fundamental constants.
\item The relevant transitions lie within the optical domain, allowing interrogation using state-of-the-art laser spectroscopy techniques.
\item It possesses long-lived excited states, enabling ultranarrow linewidths essential for high-accuracy clock performance.
\item All low-energy states of Hf$^{12+}$ are of the same parity. The states of opposite parity, which contribute to the palarizabilities are very high in the spectrum. This means that the polarizabilities are small and, as a consequence, the BBR shift is also small.
\item As in many other HCI, the clock transitions of Hf$^{12+}$ are relatively insensitive to external perturbations such as stray electric fields, motional effects in ion traps, etc.
\end{enumerate}

{ One of the clock transitions—the one between states of the ground-state configuration—was identified in our earlier work as a transition with a strongly suppressed blackbody-radiation (BBR) shift~\cite{KDF-BBR14}.
In the present study, we perform a more detailed investigation of both clock transitions using advanced computational techniques.
We calculate the electronic structure of the ion and determine its key transition properties, including energies, lifetimes, and sensitivity coefficients to variations in the fine-structure constant $\alpha$.
}

\begin{figure}
\epsfig{figure=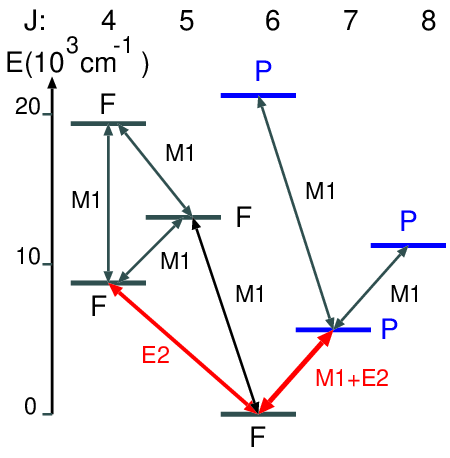,scale=0.8}
\caption{First seven energy levels of the Hf$^{12+}$ ion. The diagram is based on the data presented in Table~\ref{t:Energy}.
Possible clock transitions are shown in red as the E2 and M1+E2 transitions.
Letter F marks states of the $4f^{12}$ configuration; letter P marks states of the $4f^{11}5p$ configuration.
Corresponding values of the total angular momentum $J$ are shown on the top.}
\label{f:hf12}
\end{figure}

\section{Method of Calculation: CIPT Approach}

The ground-state configuration of Hf$^{12+}$ is [Pd]$5s^24f^{12}$. The lowest excited states belong either to this configuration or to [Pd]$5s^24f^{11}5p$ (see Fig.~\ref{f:hf12}). This ion has an open $4f$ shell, which means that all $4f$ electrons must be treated as valence electrons in the configuration interaction (CI) calculations.

The calculations begin with relativistic Hartree–Fock (RHF) computations for the ground state of the ion. These calculations show that the RHF energies of the 
$5s$ electrons are quite close to those of the $4f$ electrons. Consequently, excitations from the 
$5s$ shell may play an important role in the electronic structure. To assess this effect, we compared two approaches: in the first, the 
$5s$ electrons are treated as part of the core, while in the second they are included in the valence space. The results obtained with these two methods are very similar, at least for the low-lying states. In what follows, we adopt the approach in which the 
$5s$ electrons are treated as part of the core.

The total number of valence electrons is twelve (or fourteen if the $5s$ electrons are also included). This number is too large for standard configuration interaction (CI) calculations. To handle such systems, a specialized method known as configuration interaction with perturbation theory (CIPT) was developed in Ref.~\cite{CIPT}. The CIPT approach is based on neglecting off-diagonal matrix elements between highly excited many-electron basis states, which reduces the size of the effective CI matrix by many orders of magnitude.

The RHF Hamiltonian includes Breit interaction and quantum electrodynamic (QED) corrections,
\begin{eqnarray} \label{e:RHF}
	&&\hat H^{\rm RHF}= c\bm{\alpha}\cdot\mathbf{p}+(\beta -1)mc^2+V_{\rm nuc}(r)+ \\
	&&V_{\rm core}(r) + V_{\rm Breit}(r) + V_{\rm QED}(r), \nonumber
\end{eqnarray}
where $c$ is the speed of light, $\bm{\alpha}$ and $\beta$ are the Dirac matrices, \textbf{p} is the electron momentum, $m$ is the electron mass, $V_{\rm nuc}$ is the nuclear potential obtained by integrating the Fermi distribution of the nuclear charge density,  $V_{\rm core}(r)$ is the self-consistent RHF potential created by the electrons of the closed-shell core, $V_{\rm Breit}(r)$ is Breit potential~\cite{DzuFlaSaf06},
$V_{\rm QED}(r)$ is a model QED potential~\cite{radpot}.
	
After completing the self-consistent procedure for the core, the B-spline technique~\cite{johnson1986computation,johnson1988finite} is used to create a complete set of single-electron wave functions. The functions are constructed as linear combinations of B-splines, which are eigenstates of the RHF Hamiltonian. We use 40 B-splines of order 9 in a box with a radius of $R_{\rm max}=40a_B$; the orbital angular momentum 0~$\leq$~\textit{l}~$\leq$~6.

In the CIPT method, the CI matrix takes the form
\begin{equation}
	\langle i|H^\mathrm{eff}|j\rangle = \langle i|H^\mathrm{CI}|j\rangle+\sum_{k}\frac{\langle i|H^\mathrm{CI}|k\rangle\langle k|H^\mathrm{CI}|j\rangle}{E-E_{k}}.
	\label{e:CIPT}
\end{equation}
Here, $H^\mathrm{CI}$ is given by 
\begin{equation}
H^{\rm CI} = \sum_n H^{\rm RHF}_{{\rm core},n} + \sum_{n<m} \frac{e^2}{r_{nm}}.
	\label{e:HCI}
\end{equation}
The summation goes over valence electrons.
Indexes $i,j$ numerate low-energy states, index $k$ numerate high-energy states; $i, j \leqslant N_{\text {eff }}, N_{\text {eff }}<k \leqslant N_{\text {total}}$, 
$N_{\text {eff }}$ is the number of low-energy states and $N_{\text {total}}$ is the total number of many-electron basis states.
Note that the choice  of $N_{\text {eff }}$ is arbitrary. One can increase it until the results are stable. 
In our calculations $N_{\text {eff }} \sim 10^2$ and $N_{\text {total}} \sim 10^8$.

Parameter $E$ in (\ref{e:CIPT}) is the energy of the state of interest, and $E_k$ denotes the diagonal matrix element for high-energy states, $E_{k}=\left\langle k\left|H^{\mathrm{CI}}\right| k\right\rangle$. The summation in (\ref{e:CIPT}) runs over all high-energy states. 
Note that the parameter $E$ in the denominator of (\ref{e:CIPT}) is the same as the energy of the state of interest which is to be obtained from solving the CI equations. Since this energy is not known in advance, iterations over $E$ are needed to find it.
More detailed explanations of the technique can be found in Ref.~\cite{CIPT}.

\subsection{Calculation of transition probabilities and lifetimes}

To calculate transition amplitudes ($A$), we use the time-dependent Hartree-Fock (TDHF) method~\cite{DzuFlaSilSus87}, which is equivalent to the random-phase approximation (RPA). A TDHF equation for the core can be written as follows:
\begin{equation}\label{e:RPA}
	\left(\hat H^{\rm RHF}-\epsilon_c\right)\delta\psi_c=-\left(\hat f+\delta V^{f}_{\rm core}\right)\psi_c.
\end{equation}
Here, $ \hat H^{\rm RHF} $ is the RHF Hamiltonian (\ref{e:RHF}), index $c$ numerates states in the core, $\epsilon_c$ is the energy of electron state $c$, and $\psi_c$ is the  wave function of the state, $\delta\psi_c$ is a correction to the wave function due to the external field $\hat f$, $\delta V^{f}_{\rm core}$ is the correction to the self-consistent RHF potential of the core generated by the modifications to all core states in the external field. 

After solving the RPA equations  (\ref{e:RPA}) self-consistently for finding $\delta V^{f}_{\rm core}$, the off-diagonal matrix element then gives the transition amplitude between valence states 
\begin{equation}\label{e:Ab}
	A_{i j}=\left\langle\Psi_i\left|\hat{f}+\delta V^{f}_{\rm core }\right| \Psi_j\right\rangle
\end{equation}
Here, $\left|\Psi_i\right\rangle,\left|\Psi_j\right\rangle$ are the many-electron wave functions obtained by solving Eq.~(\ref{e:CIPT}).

In this work we calculate electric dipole amplitudes (E1) for calculating polarizabilities and BBR shift, magnetic dipole (M1) and electric quadrupole (E2) amplitudes for calculating transition probabilities between low-energy states and lifetimes of excited states.
Once the transition amplitudes have been obtained, the transition probabilities are given by (we use atomic units)
\begin{equation}\label{e:Td}
	T^{M1}_{ab} = \frac{4}{3}(\alpha\omega_{ab})^3 \frac{A^2_{M1}}{2J_b+1},
\end{equation}

\begin{equation}\label{e:Tq2}
	T^{\rm E2}_{ab} = \frac{1}{15}(\alpha\omega_{ab})^5 \frac{A^2_{E2}}{2J_b+1},
\end{equation}

In both equations, $\alpha$ is the fine structure constant ($\alpha\approx\frac{1}{137}$), $\omega_{ab}$ is the energy difference between the lower ($ a $) and upper ($ b $) states, \textit{A} is the transition amplitude (reduced matrix element) (\ref{e:Ab}), and $J_b$ is the total angular momentum of the upper state \textit{b}. Note that magnetic amplitudes $A_{M1}$ contain the Bohr magneton $\mu_B$ ($\mu_B = \alpha/2 \approx 3.65 \times 10^{-3}$ in the Gauss-type atomic units).

The lifetime (in seconds) of the upper state $b$ is given by
\begin{equation}\label{e:tau}
	\tau_b =  2.4189 \times 10^{-17}/\sum_a T_{ab},
\end{equation}
where the summation goes over all possible transitions to lower states $a$.

\section{Results and discussion}

\subsection{Energy levels, g-factors, transition amplitudes, and lifetimes}

\begin{table}[]
	\caption{\label{t:Energy}
		Excitation energies ($E$, cm$^{-1}$), $g$-factors,  and lifetimes for low-lying states of the Hf$ ^{12+} $ ion.} 
	\begin{ruledtabular}
		\begin{tabular}{rlcrcc}
			\multicolumn{1}{c}{ No. }&
			\multicolumn{1}{c}{ Conf. }&
			\multicolumn{1}{c}{  J   }&
			\multicolumn{1}{c}{ Energy  }&
			\multicolumn{1}{c}{ $g$ }&
			\multicolumn{1}{c}{ $\tau$   }\\
			\hline
 1&  $4f ^{12} $&         $ {6} $&        0.00&   1.1651   &\\
 2&   $4f ^{11} $5p&         $ {7} $&       5627 &  1.2341   &442 s\\
 3&   $4f ^{12} $&         $ {4} $&        8462  & 1.1250   &				2$\times10^4$  s\\
 4&   $4f ^{11} 5p$&         $ {8} $&       11198 &  1.1632   &\\
 5&   $4f ^{12} $&         $ {5} $&       12842 &  1.0333  &\\
 6&   $4f ^{12} $&         $ {4} $&       19422&  0.9688  &\\
 7&   $4f ^{11} 5p$&         $ {6} $&      20640 &  1.1243   &\\
 8&   $4f ^{11} 5p$&         $ {7} $&      22462  & 1.0752   &\\
 9&   $4f ^{12} $&         $ {3} $&       22755&  1.0833  &\\
10&   $4f ^{12} $&         $ {2} $&        23691 &  0.7485 &\\
11&   $4f ^{11} 5p$&         $ {5} $&      26421 &  1.0158    &\\
12&   $4f ^{11} 5p$&         $ {6} $&      28269 &  0.9953    &\\
13&   $4f ^{11} 5p$&         $ {4} $&      28353  & 1.1243    &\\
14&   $4f ^{11} 5p$&         $ {5} $&     30539 &  0.9689    &\\
15&   $4f ^{12} $&         $ {4} $&       32140  & 0.9579 &\\

\end{tabular}	
\end{ruledtabular}
\end{table}

Calculated energy levels and $g$-factors for fifteen lowest states of Hf$^{12+}$ ion are presented in Table~\ref{t:Energy}.
First seven of them are also shown on Fig.~\ref{f:hf12}. 
First two excited states of Hf$^{12+}$ ion can serve as clock states since they are connected to the ground state only by weak M1 or E2 transitions.
Corresponding calculated transition ammplitudes and rates of spontaneous decay are presented in Table~\ref{t:Tran}. 
The lifetimes of two clock states calculated via these transition using Eqs. (\ref{e:Td}), (\ref{e:Tq2}) and (\ref{e:tau}) rates are presented in Table~\ref{t:Energy}.


\begin{table*}
	\caption{\label{t:Tran} Transition amplitudes (\textit{A}, a.u.) and transition probabilities (T, 1/s) for some low  states.} 
	\begin{ruledtabular}
		\begin{tabular}{cc c cc cc }
			&&&
			\multicolumn{2}{c}{($\omega$) }&
			\multicolumn{2}{c}{Present}\\
			\cline{4-5}
			\cline{6-7}
			\multicolumn{1}{c}{Transition}& 
	 	\multicolumn{1}{c}{$J_b$}&
			\multicolumn{1}{c}{Type}&
			\multicolumn{1}{c}{ [cm$^{-1}$]}&
			\multicolumn{1}{c}{ [a.u.]}&
			\multicolumn{1}{c}{\textit{A} [a.u]}&
			\multicolumn{1}{c}{T [s$^{-1}$]}\\
			\hline			
			2-1	&7&	M1&	 5626.6&	0.0256&	-3.07E-04&	2.263E-03	\\
			2-1	&7	&E2&	 5626.6&0.0256&	0.23028	&2.231E-06\\	
			3-1&	4&	E2&	8461.8&	0.0386&	-0.30485&	5.012E-05\\	
		\end{tabular}
	\end{ruledtabular}
\end{table*}

\subsection{Cooling of the Hf$^{12+}$ ions}

Odd-parity states lie very high on the energy scale, $E > 170{,}000~\mathrm{cm}^{-1}$.
Therefore, there are no strong E1 transitions suitable for laser cooling.
An alternative approach is sympathetic cooling, in which the clock ion is co-trapped with an auxiliary “logic” ion that possesses a laser-accessible transition for cooling and state readout operations \cite{cooling1}.
The most efficient sympathetic cooling occurs when the charge-to-mass ratios ($q/m$) of the clock and logic ions are approximately equal \cite{cooling2}.
The $^9$Be$^+$ ion appears to be a suitable candidate: its $q/m$ ratio is 0.1111, compared to 0.0682 for Hf$^{12+}$.



\subsection{Polarizabilities and blackbody radiation shifts}

An important source of uncertainty in clock frequencies is blackbody radiation shift (BBR).
To estimate this shift one needs to know static scalar polarizability ($\alpha_v(0)$) for both states of the clock transition.

The polarizability of state $v$  can be expressed as a sum over a complete set of states connected to state $v$ via the electric dipole (E1) transitions 
\begin{equation}\label{e:pol}
	\alpha_v(0)=\dfrac{2}{3(2J_v+1)}\sum_{n}\frac{A_{vn}^2}{\omega_{vn}},
\end{equation}
where $J_v$ is the total angular momentum of state $v$ and $\omega_{vn}$ is the frequency of the transition. Notations $v$ and $n$ refer to many-electron atomic states. 

Expression (\ref{e:pol}) applies to valence states only. Contribution from the core is given by
\begin{equation}\label{e:polc}
\alpha_{\rm core}(0) = \frac{2}{3} \sum_c \langle \psi_c | \hat d | \delta \psi_c \rangle.
\end{equation}
Here $\hat d =-er$ is the electric dipole operator; summation goes over all core states, $\delta \psi_c$ is the correction to the core state $\psi_c$ found by solving the RPA equations (\ref{e:RPA}).
Core contribution (\ref{e:polc}) is the same for all valence states apart from small correction caused by Pauli principle: excitations from the core cannot go to the states occupied by valence electrons.
This correction is small and is not included in present calculations.

For the calculations of the polarizabilities we apply the technique developed in Ref.~\cite{Symmetry} for atoms or ions with open shells. The method relies on Eq.~(\ref{e:pol}) and the Dalgarno-Lewis approach~\cite{dalgarno1955exact}, which reduces the summation in Eq.~(\ref{e:pol}) to
solving a matrix equation (for more details, see Ref.~\cite{Symmetry}).

{ The calculated polarizabilities of the ground and excited clock states are presented in Table~\ref{t:pol}. When comparing these results with those from our earlier work~\cite{KDF-BBR14}, it should be noted that only the contributions from the $4f$ states were included in Ref.~\cite{KDF-BBR14}, since other states do not  noticeably affect the blackbody-radiation shift, which was the main focus of that study.

The polarizabilities of all three clock states are small and very similar in value (see Table~\ref{t:pol}). They are small because the transition frequencies are much lower than the excitation energies to odd-parity states ($E_{\rm odd} > 500000~{\rm cm}^{-1}$).
The similarity among them also arises from the fact that the polarizabilities are strongly dominated by the contribution of the common $5s$ state.
}
We have seen similar facts before in some other ions,
see Refs.~\cite{DF-2024,ABDFB2022,OsIr2023,ADF2024}.


In atomic clocks, the BBR shift can considerably affect the frequency of the clock's transition. 
Its value in Hz is given by
\begin{equation}
	\delta \nu_{\rm BBR} = -8.611 \times  10^{-3}
		\left(\frac{\mathrm{T}(\mathrm{K})}{300}\right)^4\Delta \alpha(0),
\end{equation}

where $T$ is temperature (room temperature is $ T $= 300 $ K $) and $ \Delta \alpha(0)= \alpha_0({\rm CS}) - \alpha_0({\rm GS})$ is the difference between the excited and ground state polarizabilities given in atomic units. 
A summary of the BBR shifts for the clock states investigated in this work is presented in Table~\ref{t:pol}. As can be seen, the relative BBR shifts for these transitions are among the smallest that have been examined so far,  $\sim 10^{-19}$. 

\begin{table*}
	\caption{\label{t:pol}
		Scalar static polarizabilities of the ground state, $\alpha_0({\rm GS})$, and clock states, $\alpha_0({\rm CS})$,  and BBR frequency shifts for the clock transitions.  $\delta\nu_{BBR}$/$\nu$ is the fractional  BBR shift; $\nu$ is the clock transition frequency. The values presented for $\alpha_0({\rm GS})$ and $\alpha_0({\rm CS})$ include core and valence contributions. The value of the core polarizability is 0.682 $a_B^3$}.
	\begin{ruledtabular}
		\begin{tabular}{cc c cr lr}
			&&&
			\multicolumn{1}{c}{$\Delta \alpha (0)$} &
			\multicolumn{3}{c}{BBR, (\textit{T}= 300 K)} \\
			\cline{4-4}
			\cline{5-7}
			
			\multicolumn{1}{c}{Transition}&
			
			\multicolumn{1}{c}{$\alpha_0({\rm GS})$[$a_B^3$]}&
			\multicolumn{1}{c}{$\alpha_0({\rm CS})$[$a_B^3$]}&
			\multicolumn{1}{c}{$\alpha_0({\rm CS}) - \alpha_0({\rm GS})$} &
			\multicolumn{1}{c}{$\delta\nu_{BBR}$[Hz]}&
			\multicolumn{1}{c}{$\nu$[Hz]}&
			\multicolumn{1}{c}{$\delta\nu_{BBR}$/$\nu$} \\
			\hline
			
			
			2 - 1  &0.7649&0.8369&0.0720&$-0.6200\times10^{-3}$&1.6869$\times10^{14}$&$-3.675\times10^{-18}$ \\
			
			3 - 1  &0.7649&0.7652&0.0003&$-0.2583\times10^{-5}$&2.5368$\times10^{14}$&$-1.018\times10^{-20}$ \\
		\end{tabular}
	\end{ruledtabular}
\end{table*}

\subsection{Quadrupole shift}


The ground and clock states of the ion under consideration have large total angular momenta ($J$=4,6, and 7). This makes them sensitive to the interaction between the gradient of the external electric field and the electric quadrupole moments of the states, which leads to a frequency shift. Therefore, accurate knowledge of the quadrupole moments is essential.

The quadrupole interaction Hamiltonian  can be written as (see, e.g., Ref.~\cite{itano2000})
\begin{equation}
	H_{\mathcal{Q}}=\sum_{q=-1}^1(-1)^q \nabla \mathcal{E}_q^{(2)} \hat{\Theta}_{-q},
\end{equation}
where $q$ is the operator component, and  the $\nabla \mathcal{E}_q^{(2)}$ tensor represents the gradient of the external electric field at the position of the system. $\hat{\Theta}_q$ is the electric quadrupole operator (the same as for the E2 transitions) and  describes as $\hat{\Theta}_q=r^2 C_q^{(2)}$, where $C_q^{(2)}$ is the normalized spherical function.

The electric-quadrupole moment $\Theta$ is defined as an expectation value of the $\hat{\Theta}_0$ operator,
\begin{equation}
	\begin{aligned}
		\Theta & =\left\langle n J J\left|\hat{\Theta}_0\right| n J J\right\rangle \\
		& =\langle n J\|\hat{\Theta}\| n J\rangle \sqrt{\frac{J(2 J-1)}{(2 J+3)(2 J+1)(J+1)}},
	\end{aligned}
\end{equation}
where $\langle n J\|\hat{\Theta}\| n J\rangle$ denotes the reduced matrix element (ME) of the electric-quadrupole operator. In Table~\ref{t:Q}, the reduced ME of the electric quadrupole operators for the considered states are shown along with their $\Theta$.
Note, that their numerical values are small, typical value for neutral atoms $\sim$ 1 a.u. This is because the operator of quadrupole moment $\propto r^2$, its matrix element is small for compact systems like HCI. Consequently, the frequency shift is also suppressed.
Further suppression can be obtained by averaging transition frequencies over measurements with different directions of applied magnetic field~\cite{Yb-clock} or constructing combinations of frequencies of the transitions between states with different projections of the total angular momentum $J$ which are not sensitive to the electric field gradient~\cite{Yb-clock,DDF-HCI12,DDF-HCI12a}.

\begin{table}[!]
	\caption{\label{t:Q}
		Quadrupole moment ($\Theta$, a.u.) of the considered optical clock states.} 
	\begin{ruledtabular}
		\begin{tabular}{crcccc}
			\multicolumn{1}{c}{No.}& 
			\multicolumn{1}{c}{Conf.}&
			\multicolumn{1}{c}{J}&
			\multicolumn{1}{c}{$E$ (cm$^{-1}$)}&
			\multicolumn{1}{c}{ ME (a.u.) }&
			\multicolumn{1}{c}{ $\Theta$ }\\&&&&
			\multicolumn{1}{c}{ $\left\langle J\|\Theta_{0}\| J\right\rangle$}&\\
			\hline
			1&   4f$ ^{12} $&         $ {6} $&        0.00&   0.4585   &0.1008\\
			2&   4f$ ^{11} $5p&         $ {7} $&       5627 & 0.0100  &0.0021\\
			3&   4f$ ^{12} $&         $ {4} $&        8462  &  -0.0057 &-0.0014\\
		\end{tabular}
	\end{ruledtabular}
\end{table}

\subsection{Sensitivity of clock transitions to the variation of the fine structure constant.}

To consider the sensitivity of atomic transitions to the variation of $\alpha$ it is conveniet to present their frequencies in the form
\begin{equation}
	\omega(x)=\omega_0+q x
\end{equation}
Here $\omega_0$ is the present laboratory value of the transition frequency and $x=\left(\alpha / \alpha_0\right)^2-1$, where $\alpha_0$ is the physical value of $\alpha$, $q$ is the sensitivity coefficient, which is to be found by varying the value of $\alpha$ in computer codes and calculating numerical derivatives
\begin{equation}
	q = \frac{\omega(+\delta)-\omega(-\delta)}{2 \delta}
\end{equation}
To ensure linear behavior, $\delta$ should be sufficiently small, yet large enough to suppress numerical noise. We used $\delta = 0.01$, which provides stable results.
To detect a possible variation of the fine-structure constant, one must measure the frequencies of at least two atomic transitions relative to each other over an extended period of time.
The ratio of the relative changes of the transition frequencies can be written as
\begin{equation} \label{e:w1w2}
	\delta\left(\frac{\omega_1}{\omega_2}\right)/\frac{\omega_1}{\omega_2}=
	\frac{\delta\omega_1}{\omega_1} - \frac{\delta\omega_2}{\omega_2} \equiv \left(K_1 - K_2 \right)\frac{\delta\alpha}{\alpha}.
\end{equation}
The parameter $K$ is known as the enhancement factor and is defined as $K=2q/\omega$.
It is evident that the highest sensitivity to a possible variation of the fine-structure constant can be achieved when two atomic transitions have significantly different sensitivities to $\alpha$ variation..
For both clock transitions studied, the values of $q$ and $K$ have been calculated and are summarized in Table~\ref{t:q}.
As seen from the table, the sensitivities of the two transitions to variations in $\alpha$ differ noticeably.
This difference arises because the $ 2 \to 1$ transition occurs between states of different electronic configurations — specifically, it corresponds to a 
$p \to f$ single-electron transition. In contrast, the  $ 3 \to 1$ transition takes place between states of the same configuration.
For these two transitions, Eq.~(\ref{e:w1w2}) takes the form

\begin{equation} \label{e:w11w22}
\frac{\delta (\omega_1/\omega_2)}{(\omega_1/\omega_2)} = -69 \frac{\delta\alpha}{\alpha}.
\end{equation}
For comparison, one of the highest sensitivities to $\alpha$ variation realized in current optical clocks is obtained from the frequency ratio of the E2 and E3 transitions in Yb$^+$, where $\Delta K \approx 7$  \cite{FlaDzuCJP}.
The sensitivity of clock transitions in Hf$^{12+}$ to variation of $\alpha$ is similar to recently proposed transtions in  Cf$^{16+}$,  and Cf$^{17+}$~\cite{BDFO-Cf12}. The advantage of Hf is that it is stable element. 

Note that having large enhancement factor $K$ is not the only consideration in searching for good atomic systems. High precision  of the measurements is equally important (see, e.g. Ref.~\cite{Yb-clock} for a detailed discussion). Assuming that the uncertainty of the measurements of atomic frequency ($\Delta \omega$) obtained over some period of time serves as a limit on the time-variation of $\alpha$, one can get from (\ref{e:w1w2})
\begin{equation}\label{e:dalpha}
\frac{\delta \alpha}{\alpha} < \frac{\Delta \omega}{2q_1}.
\end{equation}
Here $\Delta \omega = \sqrt{\Delta \omega_1^2 +  \Delta \omega_2^2}$, $q_2$ is neglected because $q_1 \gg q_2$.
This implies that to obtain the strongest limits on the time variation of $\alpha$, one should search for atomic clock transitions that combine high measurement precision (small $\Delta \omega$) with high sensitivity to $\alpha$ variation (large $q$).
In particular, when seeking a large enhancement factor $K,$ it is more effective to look for transitions with large 
$q$ rather than those with small transition frequencies.


\begin{table}[!]
	\caption{\label{t:q}
		Sensitivity of the clock transitions to the variation of the fine-structure constant ($q, K$) of Hf$^{12+}$.} 
	\begin{ruledtabular}
		\begin{tabular}{cccccc}
						
			\multicolumn{1}{c}{Transition}&
			\multicolumn{1}{c}{Conf.}&
			\multicolumn{1}{c}{J}&
						
			\multicolumn{1}{c}{$\omega$ (cm$^{-1}$)}&
			
			\multicolumn{1}{c}{$q$(cm$^{-1}$) }&
			\multicolumn{1}{c}{$K$}\\
			\hline	
			
		2 - 1&   4f$ ^{11} $5p&         $ {7} $&       5627 &  -195221 &-69.4\\		
		3 - 1&   4f$ ^{12} $&         $ {4} $&        8462  & -1261  &-0.30\\
					
		\end{tabular}
	\end{ruledtabular}
\end{table}


\section{Conclusion}

Two clock transitions in the Hf$^{12+}$ ion have been studied theoretically. The first is the transition between the ground state 
$4f^{12} \ (J=6)$ and the metastable state $4f^{11}5p\ (J=7)$.
This is an $f-p$ transition that exhibits high sensitivity to variation of the fine-structure constant $\alpha$
The second is an electric-quadrupole (E2) transition between states of the ground-state configuration. It is largely insensitive to variations in 
$\alpha$ and can serve as an anchor transition when one clock frequency is measured against the other.

Both transitions are highly insensitive to external perturbations. In particular, they feature an extremely small blackbody-radiation (BBR) shift, owing to the fact that the static dipole polarizabilities of all three relevant states are small and nearly equal. The quadrupole and other systematic shifts are also small, as is typical for highly charged ions (HCI). The Hf$^{12+}$ ion can be sympathetically cooled by co-trapping with Be$^+$ ions.

Our study demonstrates that Hf$^{12+}$ is a promising candidate for both precision timekeeping and tests of fundamental physics.


\bibliographystyle{apsrev4-2}
\bibliography{hf12,dzuba}

\end{document}